\documentclass[aps,prb,twocolumn,groupedaddress]{revtex4}
\newcommand{\ce}{CeCu$_2$Si$_2$}
\newcommand{\cege}{CeCu$_2$Ge$_2$}

\usepackage{multirow}
\usepackage{graphicx}
\begin{document}

\title{Signatures of valence fluctuations in \ce~under high pressure}
\author{Alexander T. Holmes}
\email{alexander.holmes@physics.unige.ch}
\author{Didier Jaccard}
\email{didier.jaccard@physics.unige.ch}
\affiliation{D\'{e}partement de Physique de la Mati\`{e}re
Condens\'{e}e, Section de Physique, University of Geneva, 24 quai
Ernest-Ansermet CH-1211 Gen\`{e}ve 4, Switzerland}
\author{Kazumasa Miyake}
\email{miyake@mp.es.osaka-u.ac.jp}
\affiliation{Division of
Materials Physics, Department of Physical Science, Graduate School
of Engineering Science, Osaka University, Toyonaka, Osaka
560-8531, Japan}
%\affiliation{}
\date{\today}
\begin{abstract}
Simultaneous resistivity and a.c.-specific heat measurements have
been performed under pressure on single crystalline \ce~to over 6
GPa in a hydrostatic helium pressure medium. A series of anomalies
were observed around the pressure coinciding with a maximum in the
superconducting critical temperature, $T_c^{\rm max}$. These
anomalies can be linked with an abrupt change of the Ce valence,
and suggest a second quantum critical point at a pressure $P_{\rm
v}\simeq 4.5\:$GPa, where critical valence fluctuations provide
the superconducting pairing mechanism, as opposed to spin
fluctuations at ambient pressure.  Such a valence instability, and
associated superconductivity, is predicted by an extended Anderson
lattice model with Coulomb repulsion between the conduction and
f-electrons. We explain the $T$-linear resistivity found at
$P_{\rm v}$ in this picture, while other anomalies found around
$P_{\rm v}$ can be qualitatively understood using the same model.
\end{abstract}
\maketitle
\section{Introduction}\label{sec:Introduction}
Boosted by the discovery of superconductivity in \ce~over twenty
years ago,\cite{Steglich79} the relationship between
superconductivity and magnetism has been extensively investigated
in various d and f heavy fermion (HF) compounds. A consensus has
developed that HF superconductivity is mediated by spin
fluctuations,\cite{Julian98,MSRV,Scalapino,Monthoux02} mainly
because superconductivity was found close to a magnetic
instability at $T=0$, sometimes described as a quantum-critical
point (QCP), often attained by applying pressure. A recent
development is that an essentially gapless superconducting (SC)
state has been identified by NMR/NQR measurements in the region
where the SC state coexists with
antiferromagnetism,\cite{KawasakiY01, KawasakiS02} consistent with
a theoretical prediction.\cite{Fuseya03}
%\\

In this paper we further explore the possibility that at a
pressure $P_{\rm v}\simeq 4.5\:$GPa, a second QCP, associated with
f-electron occupation number, has a major role to play in the
superconductivity of \ce~and related
compounds.\cite{Miyake99,Jaccard99}
%\\

\ce\ has a superconducting ground state at ambient pressure with a
critical temperature $T_c$, around 0.7~K.  It is firmly believed
that the compound is close to an antiferromagnetic QCP at slight
negative pressure, accessible for example by partial substitution
of Si with Ge.\cite{Yuan03} When pressure is applied, $T_c$
initially remains close to its ambient pressure value, followed by
a sudden increase to around $2\:$K at about $3\:$GPa. Further
increase in pressure results in a slower suppression of $T_c$ to
zero. This non-monotonous behavior of $T_c(P)$ was first explored
by resistivity in the quasi-hydrostatic conditions of Bridgman
anvil cell.\cite{Bellarbi84} Subsequent investigations by
susceptibility\cite{Thomas96} and resistivity
\cite{Vargoz98,Thomasson98} were carried out in various pressure
media, and showed considerable variation in $T_c$ between samples,
especially at high pressure.
%\\

With increasing pressure \ce~passes from a nearly trivalent $4\rm
f^1$ behavior, with Kondo coupling between conduction and
f-electrons, to behavior at very high pressure characteristic of
intermediate valence (IV) systems, whose valence fluctuates
between the $4\textrm{f}^n$ and $4\textrm{f}^{n-1}+[5\rm d6\rm s]$
electronic configurations. As a result, deep in this IV regime,
the resistivity, for instance, resembles that of LaCu$_2$Si$_2$,
which lacks $4\rm f$ electrons. Similar $T_c(P)$ dependence to
that found in \ce~is seen in the isoelectronic sister compound,
\cege, offset by about 10$\:$GPa due to the larger atomic volume
of Ge.\cite{Vargoz98b} Apart from this shift of the pressure
scale, no qualitative difference occurs between the two compounds
beyond unavoidable sample dependence.

From a more theoretical point of view, there exist at least three
reasons to believe that critical valence fluctuations are at the
origin of the pressure-induced peak of the SC transition
temperature $T_c$.
%\\

First, the $A$ coefficient of the $T^{2}$ resistivity law
decreases drastically by about two orders of magnitude around the
pressure corresponding to the $T_c$ peak.\cite{Jaccard99} Since
$A$ scales as $(m^{*}/m)^{2}$ in the so-called Kondo regime, this
implies that the effective mass $m^{*}$ of the quasiparticles also
decreases sharply there.  This fall of $m^{*}$ is possible only if
there is a sharp change of Ce valence, deviating from Ce$^{3+}$,
since the following approximate formula for the renormalization
factor $q$ holds in the strongly correlated
limit:\cite{RiceUeda,Shiba}
\begin{equation}\label{eq:m*nf}
{m^{*}\over m} \simeq q^{-1}={1-n_{\rm f}/2\over 1-n_{\rm f}},
\end{equation}
where $n_{\rm f}$ is the f-electron number per Ce ion.
%\\

Second, the so-called Kadowaki-Woods (KW) ratio,\cite{Kadowaki86}
$A/\gamma^{2}$, where $\gamma$ is the Sommerfeld coefficient of
the electronic specific heat, crosses over quickly from that of a
strongly correlated class to a weakly correlated one.\cite{MMV}
The inverse of the Sommerfeld coefficient, $\gamma^{-1}$, can be
identified with the Kondo temperature $T_K$, which is
experimentally accessible by resistivity measurements. This
indicates that the mass enhancement due to the dynamical electron
correlation is quickly lost at around $P\sim P_{\rm v}$, in
agreement with the previous point. The phenomenon can be
understood if we note the fact that $\gamma$ consists essentially
of two terms:
\begin{eqnarray}
\gamma& &=\gamma_{\rm band}\biggl
(1-{\partial\Sigma(\epsilon)\over\partial\epsilon}\biggr),
\nonumber
\\
& &\equiv\gamma_{\rm band}+\gamma_{\rm cor},
\end{eqnarray}
where $\gamma_{\rm band}$ is due to the so-called band effect and
$\gamma_{\rm cor}\equiv-\gamma_{\rm band}
\partial\Sigma(\epsilon)/\partial\epsilon$ is
due to the many-body correlation effect, with $\Sigma(\epsilon)$
being the self-energy of the correlated electrons. $\gamma_{\rm
cor}$ and
 $A$ are related to each other through the
Kramers-Kr\"onig relation, leading to the large value of the KW
ratio,\cite{MMV} and when $\gamma_{\rm cor}\gg\gamma_{\rm band}$,
this is indeed seen. On the other hand, if $\gamma_{\rm
cor}\sim\gamma_{\rm band}$, the ratio $A/\gamma^{2}$ should be
reduced from the KW value considerably because the effect of
$\gamma_{\rm band}$ cannot be neglected in its denominator.
%\\

Third, there is a sharp peak in the residual resistivity
$\rho_{0}$ at around $P\simeq P_{\rm v}$,\cite{Jaccard99} which
can be understood as a many-body effect enhancing the impurity
potential (in fact we define the pressure $P_{\rm v}$
experimentally by the maximum of $\rho_0$). In the forward
scattering limit, this enhancement is proportional to the valence
susceptibility $-(\partial n_{\rm f}/\partial \epsilon_{\rm
f})_{\mu}$, where $\epsilon_{\rm f}$ is the atomic f-level of the
Ce ion, and $\mu$ is the chemical potential.\cite{Miyake02}
Physically speaking, local valence change coupled to the impurity
or disorder gives rise to a change of valence in a wide region
around the impurity which then scatters the quasiparticles quite
strongly, leading to the increase of $\rho_{0}$. The enhancement
of $\rho_{0}$ can be thus directly related to the degree of
sharpness of the valence change, because the variation of the
atomic level $\epsilon_{\rm f}$ is considered to be a smooth
function of the pressure.
%\\

These circumstantial clues to the importance of critical valence
fluctuations have been backed up by a microscopic calculation of
$T_c$ for $d$-wave pairing as a function of $\epsilon_{\rm
f}$.\cite{Onishi00} This showed that sudden valence change occurs
if a moderately sized Coulomb repulsion $U_{\rm cf}$ is taken into
account between the conduction c-, and localized f-electrons, with
the peak structure of $T_c$ being qualitatively reproduced.
\begin{table*}%[H] add [H] placement to break table across pages
\begin{ruledtabular}
\begin{tabular}{p{20pt}l|cc}
    &           & \ce&\cege\\%CePd$_2$(Si/Ge)
    \hline
    \multirow{4}{*}{\textbf{(i)}}&Volume discontinuity&-&\onlinecite{Onodera02}\\
    &$L_{III}$ X-ray absorption&\onlinecite{Roehler88}&-\\
    &Drastic change of $A$ by two orders of magnitude&This work,
\onlinecite{Jaccard99}&\onlinecite{Jaccard99} \\
     &Change of $A \propto(T_1^{\rm{max}})^{-2}$ scaling &This work,
\onlinecite{Jaccard99}&\onlinecite{Jaccard99} \\
    \hline
    \multirow{2}{*}{\textbf{(ii)}} &Maximum in $T_c(P)$&This work,
\onlinecite{Bellarbi84}&\onlinecite{Vargoz98b}\\
    &Large peak in $\rho_0$          &This work,\onlinecite{Jaccard99}
&\onlinecite{Jaccard99} \\%
    \hline
     \multirow{2}{*}{\textbf{(iii)}}&Maximum in $\gamma \simeq (C_P/T)$&This
work,\onlinecite{Vargoz98}&-\\%
    &$\rho\propto T^n$ from $T_c<T<T^*$, with $n(P_{\rm v})=1$ minimum&    This
work,\onlinecite{Bellarbi84,Jaccard85}&\onlinecite{Jaccard99}  \\
   \hline
     \multirow{4}{*}{\textbf{(iv)}}    &Sample dependence of $T_c$ & This
work,\onlinecite{Bellarbi84,Thomas96,Thomasson98,Jaccard98,Jaccard99,VargozPhD}
&\onlinecite{Jaccard99} \\
     &Enhanced  $\left.\frac{\Delta C_P}{\gamma T}\right|_{T_c}$&This work&-\\
    &Resistivity and thermopower indicate $T_1^{\rm{max}}\simeq T_2^{\rm{max}}
$&\onlinecite{Jaccard99,Jaccard85}&\onlinecite{Jaccard99,Link96}  \\
    &Broad superconducting transition widths $\Delta T_c$ &This work,
\onlinecite{Bellarbi84}&\onlinecite{Jaccard99} \\
\end{tabular}
\caption{\label{tab:properties} Anomalies in \ce\ and \cege\
associated with valence
transition, with references. Symbols explained in the text.\\
Part \textbf{(i)}: Direct evidence for sudden valence change.\\
Part \textbf{(ii)}: Anomalies explained by published
valence fluctuation theory.\cite{Onishi00,Miyake02}\\
Part \textbf{(iii)}: Anomalies explained by extended treatment of
the critical valence fluctuations
(section \ref{sec:theory}).\\
Part \textbf{(iv)}: Other anomalies observed around crossover to
intermediate valence with pressure.}
\end{ruledtabular}
\end{table*}

Table \ref{tab:properties} summarizes the current experimental
evidence of anomalies seen in CeCu$_2$(Ge/Si)$_2$ around $P_{\rm
v}$.
%\\

Part \textbf{(i)} of table \ref{tab:properties} refers to direct
evidence for a valence transition of the Ce ion: Cell
volume\cite{Onodera02} and $L_{III}$ X-ray
absorption\cite{Roehler88} measurements show discontinuities as a
function of pressure. The drastic decrease of the $A$ coefficient
of the $T^2$ resistivity law, along with the $A$ vs
$T_1^{\rm{max}}$ scaling relation, indicate that the system is
leaving the strongly correlated regime characterized by a
f-occupation number close to unity. ($T_1^{\rm{max}}$ is defined
in Fig.~\ref{fig:Tmax} and assumed to be proportional to $T_K$.)
%\\

Part \textbf{(ii)} refers to anomalies observed close to the
maximum of $T_c$ predicted by critical valence fluctuation
theory.\cite{Onishi00,Miyake02} These are the maximum of $T_c$
itself and the enhanced residual resistivity,
$\rho_0$.\cite{Jaccard99,Thomas96,Bellarbi84,Vargoz98,Thomasson98,Vargoz98b}
%\\

Part \textbf{(iii)} refers to properties following from the
extended treatment of the critical valence fluctuations found in
section \ref{sec:theory} of this paper. This includes the linear
resistivity,\cite{Jaccard99} and the maximum in $\gamma$, both
found around $P_{\rm v}$.
%\\

In part \textbf{(iv)} are listed the remaining features that are
observed in \ce\ and \cege\ around the maximum in $T_c$ but which
are so far not fully explained. For example the merging of
$T_1^{\rm{max}}$ and $T_2^{\rm{max}}$, where the latter (also
defined in Fig.~\ref{fig:Tmax}) is believed to reflect the effect
of the excited crystalline electric field (CEF) split f-levels.
Many of the anomalies noted in table \ref{tab:properties} have
also been observed to coincide with the maximum of $T_c$ in other
HF superconductors, from CePd$_2$Si$_2$\cite{Demuer02} to
CeCu$_5$Au\cite{Wilhelm00}, the latter showing traces of
superconductivity under pressure.
%\\

Previous work on \ce~has shown a lot of variation in
low-temperature behavior between different samples. The fact that
not all reports have shown every anomaly is not entirely
surprising, since large variations in the electronic properties of
\ce~are well known to result from extremely small differences in
composition.\cite{Ishikawa83,Steglich96} The extension of these
variations with pressure has not been systematically explored, but
almost all samples so far studied have shown an enhancement of
$T_c$, along with effects such as the enhancement of the residual
resistivity, to be discussed below.
%\\

This variability under pressure may be due to the samples
themselves, or to pressure inhomogeneities caused by
non-hydrostatic pressure media. We were therefore motivated to use
solid helium as a pressure medium, due to its near-ideal
hydrostaticity at low temperature. By simultaneously probing
resistivity and specific heat in the same sample, we were able to
explore both percolative transport and bulk evidence for
superconductivity.
\section{Experimental methods}
\label{sec:methods} High pressure was induced using diamond anvils
with a 1.5$\:$mm culet.\cite{Eremets96} A stainless steel gasket
was specially prepared to absorb the large volume decrease of the
helium pressure medium from ambient pressure, and to avoid
severing the measurement wires.  These were insulated from the
gasket using a mixture of Al$_2$O$_3$ powder and epoxy resin.  The
pressure was measured to within 0.02$\:$GPa at various
temperatures down to 4.2$\:$K using the ruby fluorescence scale.
%\\

The \ce~sample was prepared by reaction of its constituent
elements with a slight excess of Cu, with a nominal initial
composition CeCu$_{2.1}$Si$_2$. The product was then melted in an
induction furnace and slowly allowed to crystallize under 50 bars
Ar in a BaZrO$_3$ crucible (see Ref.~\onlinecite{Vargoz98} for
more details).
%\\

The small monocrystal used in this work was cut and polished to
$230\times 80\times 20\:\mu{}$m$^{3}$, and six 5$\:\mu$m $\phi$
wires (four gold and two Au + 0.07\% at.\ Fe) were spot welded to
the sample. The $c$-axis of the tetragonal structure was parallel
to its smallest dimension. The magnetic field, when applied, was
parallel to the $c$-axis.
%\\

The six wires spot-welded to the sample allowed multiple redundant
measurements to be performed. This improved reliability and
enabled us to verify the calorimetry measurements using several
different configurations. The sample resistance could be measured
by a four-point method; a knowledge of the sample dimensions then
enabled the absolute resistivity to be determined to within 10\%.
%\\

The two thermocouple junctions were formed from an
Au/\underline{Au}Fe pair at either end of the sample. An
alternating resistive heating current was passed through one (to
avoid passing the current through the sample), while the signal
from the other was measured using a lock-in amplifier. The
resulting temperature oscillations serve as a sensitive measure of
the sample heat capacity.\cite{Eichler79} A simple model of the
a.c.-calorimetry system predicts the amplitude and phase of the
temperature oscillations ($T_{AC}$) induced by a.c. heating:
\begin{equation}\label{Tac}
T_{AC} = \frac{P_0}{K+i \omega C}\ ,
\end{equation} where $P_0$ is the
heating power, $K$ the thermal conductance to the bath, $C$ the
heat capacity, and $\omega/2 \pi$ the excitation frequency,
assumed to be low enough that the thermometer can follow the
temperature oscillations (the factor of $2\pi$ may be assumed
implicitly from this point). The signal therefore contains a
contribution from the specific heat and from thermal coupling to
the surroundings.
%\\

For frequencies $\omega\gg\omega_c$, where $\omega_c$ is the
cut-off frequency $K/C$,  the sample contribution dominates the
signal, and $\left|T_{AC}\right|$ can be considered to be
inversely proportional to the heat capacity (which we assume to be
dominated by the sample). For $\omega\ll\omega_c$, the signal
approaches the d.c.~limit and gives a measure of the mean
elevation of the sample temperature over that of the bath. For
intermediate measuring frequencies, information from the phase
$\theta$, can be used to extract the specific heat:
\begin{equation}\label{eq:phase} C = \frac{-P_0 \sin \theta}{\omega
\left|T_{AC}\right|}\ .
\end{equation} Alternatively, one can
subtract a background signal taken at a different frequency, with
\begin{equation}\label{eq:Tdc}
    C = \frac{P_0}{(\omega_2^2-\omega_1^2)^{1/2}}
    \left(\frac{1}{\left|T_{AC,\omega_2}\right|^2}-\frac{1}{\left|T_{AC,\omega_1}\right|^2}\right)^{1/2},
\end{equation} where ideally $\omega_2>\omega_c>\omega_1$.
The sample temperature must also be corrected for the constant
d.c.\ component of the oscillatory Joule heating. This was done by
repeating the measurement well below the cut-off frequency, also
providing the background signal in order to estimate $C_P$ using
eq.$\:$(\ref{eq:Tdc}).
%\\

The cut-off frequency $\omega_c$ turned out to be very temperature
dependent, varying between 200$\:$Hz at 0.5$\:$K and over 2$\:$kHz
at 1.5$\:$K, presumably due to the thermal properties of the
surrounding material. Fortunately, while complicating the data
analysis, the reduction in $\omega_c$ at the lowest temperatures
allows the technique to be used down to $\sim$100$\:$ mK. The two
estimates of $C_P$ using eqs.~(\ref{eq:phase}) and (\ref{eq:Tdc})
are in good agreement below $\sim$2$\:$K.  The working frequency
$\omega_2$ was generally of the order of $\omega_c$.
%\\

Sources of systematic error in the result might come from:
variation of the \underline{Au}Fe thermopower under pressure;
temperature and/or frequency dependent addenda to the measured
specific heat due to the pressure medium, gasket and/or anvils; or
any irreversibility or first-order character in the transitions
being observed. These potential problems will be addressed in the
discussion.
\section{Experimental results}\label{sec:expt}
We present five principal results from the sample reported in this
paper, and by drawing on previous work, we aim to place our work
in a broader context. We will try to highlight common features
found in many samples of \ce, one of the defining characteristics
of which is its variability.

\begin{itemize}
    \item We present the superconducting phase diagram obtained
using various criteria for $T_c$, and compare it to the widely
quoted phase diagram determined under hydrostatic conditions by
susceptibility.
    \item We examine the details of the superconducting transition,
    which provides some insight into the nature of the
SC state and into the sample itself.
    \item We estimate the variation of
the Sommerfeld coefficient $\gamma$, with pressure, and compare it
to previous results obtained by analysis of the upper critical
field.
    \item We report the pressure dependence of the residual
resistivity $\rho_0$, and exponent $n$, determined by a fit to the
normal state resistivity of $\rho=\rho_0+\tilde{A}T^n$ ($\tilde A$
denoting a free exponent as opposed to the quadratic coefficient
$A$). A comparison of $\rho_0(P)$ between different samples
reveals a scaling relation which can be related to the theoretical
enhancement of impurity scattering.
    \item We explore the deviation from the scaling relation
$A\propto T_K^{-2}$, which indicates the sharp change in
f-electron occupation number described in the introduction. The
enhancement of $T_c$ and the other results described above are
shown to occur around the same pressure.
\end{itemize}

\begin{figure}
    \includegraphics[width=\columnwidth]{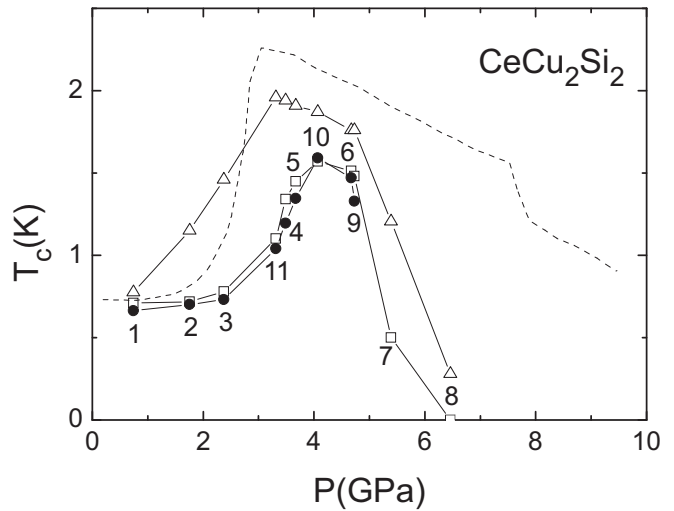}%
\caption{\label{fig:Tc} $T_c(P)$ in CeCu$_2$Si$_2$ determined from
resistivity and specific heat measurements. The triangles show
$T_c$ determined from the onset of the resistive transition
($T_c^{\rm{onset}}$), the squares show its completion
($T_c^{R=0}$), and the filled circles show the midpoint of the
specific heat jump. The numbers indicate the sequence of
pressures. The dotted line shows $T_c$ determined by
susceptibility in a different sample, also in a helium pressure
medium.\cite{Thomas96}}
\end{figure}

Figure~\ref{fig:Tc} shows the superconducting phase diagram
determined by both resistivity and specific heat, both on
increasing and decreasing the pressure. Two qualitatively
different types of behavior can be seen in the same sample,
represented by the onset and completion of the resistive
transition.
%\\

If we follow the transition onset $T_{c}^{\rm{onset} }(P)$, one
sees the sharp kinks similar to those seen in
Ref.~\onlinecite{Thomas96} (dashed line), along with a linear
decrease of $T_c$ between 3.3 and 4.8$\:$GPa at a rate of
0.14$\:$K$\,$GPa$^{-1}$. Superconductivity is observed however
over a much smaller pressure range in our sample than in
Ref.~\onlinecite{Thomas96}.
%\\

The temperature $T_c^{R=0}(P)$, at which the resistance vanishes,
behaves differently from $T_{c}^{\rm{onset}}(P)$. It has a
narrower peak with a maximum at slightly higher pressure.
$T_c^{R=0}$ agrees closely however with the transition seen in the
specific heat (see below). When a magnetic field was applied,
$T_c^{R=0}$ and the specific heat anomaly shifted in agreement.
%\\

The large resistive transition widths found in \ce~at high
pressure are often blamed on a lack of hydrostaticity due to the
pressure medium. As helium was used in this case, we can rule out
pressure inhomogeneities and concentrate on the sample itself.
Further information about the SC state comes from the effect of
measurement current on the transition width. For example, at
1.78$\:$GPa high current led to the upper part of the transition
disappearing, and a resistive transition can even be recovered
with a narrow width comparable to that close to ambient pressure.
This is presumably due to the presence of filamentary
superconductivity, with a higher $T_c$, whose critical current
density is exceeded. These broad resistive transitions appear to
be a universal feature of \ce\ at high pressure. Let us recall
that even for the highest $T_{c}^{\rm{onset} }$ measured in a
single crystal, at 2.4$\:$K, a tail of 1\% of the normal state
resistivity remained well below 2$\:$K, vanishing only at
1.5$\:$K.\cite{Vargoz98} The status of the superconductivity of
\ce~between $T_{c}^{\rm{onset}}$ and $T_c^{R=0}$ remains
mysterious.
%\\
\begin{figure}
    \includegraphics[width=\columnwidth]{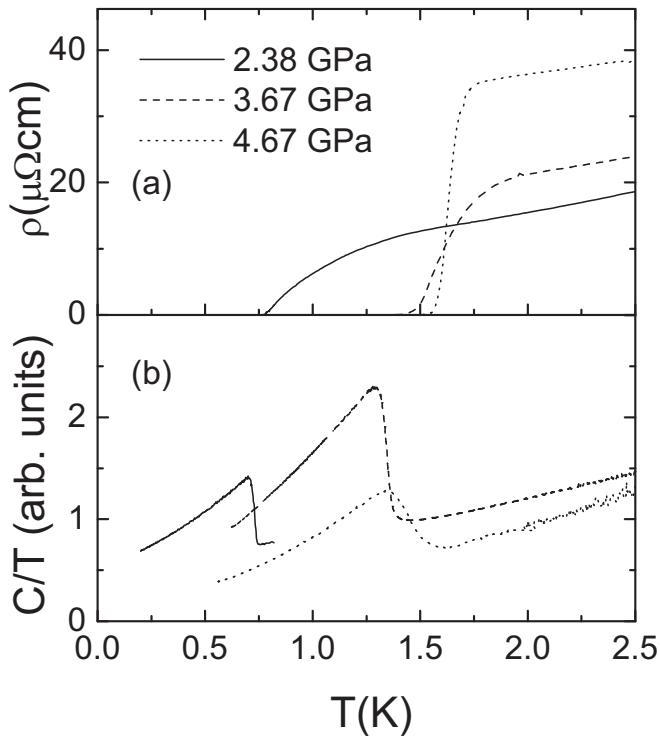}
\caption{\label{fig:Cp} Superconducting transition at three
pressures in (a) resistivity and (b) specific heat.  Note the
width of the resistive transitions, and the fact that the start of
the jump in specific heat coincides with the completion of the
resistive transition.}
\end{figure}

Figure~\ref{fig:Cp} compares the superconducting transition in
resistivity and specific heat at three different pressures. At
2.38$\:$GPa the resistive transition is broad and the sharp
specific heat jump at 0.73$\:$K begins at the point where the
resistance falls to zero.  At 3.67$\:$GPa the specific heat jump,
at 1.35$\:$K, is much larger, and remains sharp (and did so at
intervening pressures), while the corresponding resistive
transition has narrowed considerably. At 4.07$\:$GPa (not shown)
where $T_c^{R=0}$ has a maximum around 1.6$\:$K in both $\rho$ and
$C_P$, the specific heat peak has already started to broaden and
collapse in amplitude, while at the same pressure the resistive
transition is at its narrowest since ambient pressure. As $T_c$ is
driven to zero at high pressure, the superconducting $C_P$ jump
becomes smaller and broader (as shown at 4.67$\:$GPa) until it is
no longer visible. When the pressure was reduced, the $C_P$ peak
recovered its shape, indicating the reversibility of the bulk
pressure-induced behavior.
%\\

 The dramatic increase in the apparent size of the
superconducting jump is intriguing, and might suggest the presence
of strong coupling,\cite{Tsuneto98} or other qualitative change in
the SC state. Although the apparent value of $\left(\Delta
C_P/\gamma T\right)_{T_c}$ is clearly less than the BCS ratio of
1.43, similar a.c. measurements on CeCoIn$_5$ in an argon pressure
medium indicate that there is a substantial contribution to the
measured heat capacity from addenda.\cite{braithwaitepc} In helium
we would expect this to be even more significant

The increase in the $C_P$ jump size might itself be an artefact of
the uncalibrated a.c.-calorimetry method; nevertheless
$\left(\Delta C_P/\gamma T\right)_{T_c}$ does appear to show a
maximum at a pressure coinciding with the increase in $T_c$.
Furthermore, the assumption of strong coupling provided the best
fit to $H_{c2}$ for measurements of the upper critical field in
another sample.\cite{Vargoz98}

\begin{figure}
    \includegraphics[width=\columnwidth]{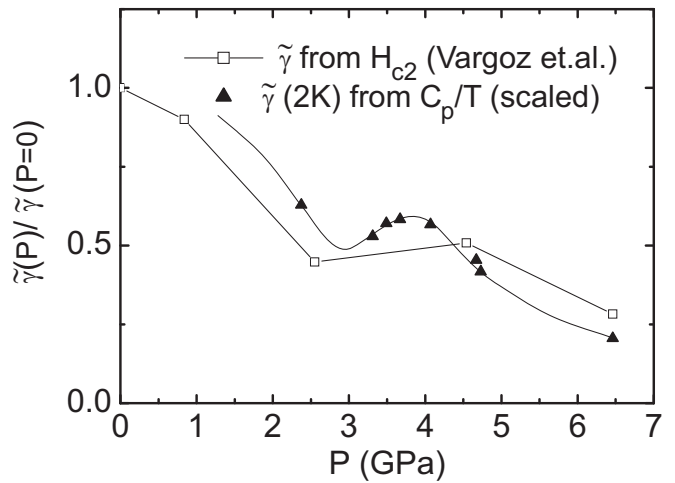}%
\caption{\label{fig:gamma}Estimate $\tilde\gamma(P)$ of the
Sommerfeld coefficient from a.c.-calorimetry signal at 2$\:$K
(triangles), scaled for comparison with that deduced from $H_{c2}$
measurements (squares).\cite{Vargoz98} The lines are guides for
the eye.}
\end{figure}
%\\

The electronic specific heat coefficient $\gamma$, and hence the
effective mass $m^*/m$, can be estimated by following the
calorimetric signal $C/T$, at a fixed temperature and measurement
frequency above the superconducting transition, though this
includes constant or slowly varying addenda from the helium,
diamonds etc. Figure~\ref{fig:gamma} shows the estimate
$\tilde\gamma(P)$, along with the value deduced from measurements
of the upper critical field in Ref.~\onlinecite{Vargoz98}. A
single constant scale factor has been introduced, showing that the
two curves can be superimposed. There is a clear anomaly in
$\tilde\gamma$ at 4$\:$GPa (just below the pressure corresponding
to $T_c^{\rm{max}}$), superimposed on a constant reduction with
pressure. The effective mass is also reflected in the initial
slope of the upper critical field $H_{c2}'(T_c)$, which in our
sample also had a maximum at the same pressure as the peak in
$\tilde\gamma$.
\begin{figure}
    \includegraphics[width=\columnwidth]{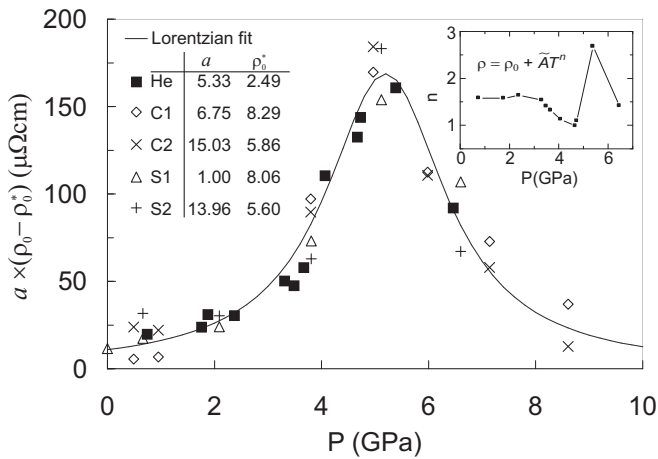}
\caption{\label{fig:rho0}Enhancement of residual resistivity in
several different \ce~samples, scaled to a universal pressure
dependence, with $a$ and $\rho_0^*$ being normalizing factors. The
maximum in $\rho_0$ is at a pressure slightly higher than that
corresponding to the maximum in $T_c$.  The inset shows $n$ for a
fit of $\rho=\rho_0+\tilde{A}T^n$. Filled squares indicate results
from this work, the rest are reported in
Ref.~\onlinecite{VargozPhD} (S and C refer to the original labels
and a retained for continuity).}
\end{figure}
%\\

The residual resistivity $\rho_0$, has a huge peak at a pressure
slightly higher than the maximum in $T_c$.  The magnitude of this
peak varies by a factor of more than ten between
samples.\cite{VargozPhD,Vargoz98} However, it is possible to scale
the residual resistivities from different samples onto the same
lorentzian curve (see Fig.\ \ref{fig:rho0}). A constant value
$\rho_0^*$, different for each sample, is subtracted from $\rho_0$
at each pressure, and the result is multiplied by a scale factor,
$a$ (i.e. $\rho_0'=a(\rho_0-\rho_o^*)$), so that all lie on the
curve defined by sample S1, which has the highest residual
resistivity (i.e. $a_{\text{S1}}=1$).
%\\

According to the theoretical prediction, the residual resistivity
$\rho_{0}$ is given as \cite{Miyake02}
\begin{equation}
\rho_{0}=Bn_{\rm imp}|u(0)|^{2}\ln\biggl|\biggl( -{\partial n_{\rm
f}\over\partial\epsilon_{\rm f}}\biggr)_{\mu}/N_{\rm F}\biggr|
+\rho_{0}^{\rm unit},
\end{equation}
where the coefficient $B$ depends on the band structure of host
metals, $n_{\rm imp}$ is the concentration of impurities with
moderate scattering potential $u(q)$ coming from disorder other
than Ce ions, $N_{\rm F}$ is the density of states of
quasiparticles around the Fermi level, and the last term
represents the residual resistivity due to unitary scattering
mainly arising from any deficit or defect of the Ce ions. The
scaling behavior of $\rho_{0}$ shown in Fig.~\ref{fig:rho0} would
be possible if the universal form is given by $\ln|(-\partial
n_{\rm f}/\partial\epsilon_{\rm f})_{\mu}/N_{\rm F}|$. It is an
open question whether the observed lorentzian form is indeed
reproduced by the theory of Ref.~\onlinecite{Onishi00}.
%\\

There is a striking correlation between the scaling factor $a$ and
the behavior of $T_c$. The sample measured in helium reported in
this paper and sample C1, pressurized in steatite, both have
similar values of $a$, and both $T_c^{\text{\rm{onset} }}$ and
$T_c^{R=0}$ agree over almost the entire pressure range. Sample
S1, with the highest $\rho_0$ at $P_{\rm v}$, has a lower
$T_c^{\rm max}$ ($\simeq1.2\:$K), and the superconductivity
disappears at a lower pressure. Samples C2 and S2 have scaling
factors $a$ around 14, and show a higher maximum $T_c$, with
superconductivity extended over a greater pressure range than in
the samples with larger residual resistivities. These differences
between samples, both in $\rho_0$ and $T_c$, are vastly amplified
from their appearance at ambient pressure.
%\\

The inset in Fig.~\ref{fig:rho0} shows the result of a fit to
$\rho=\rho_0+\tilde{A}T^n$ above $T_c$. There are two important
points to note here. Firstly, at the pressure slightly higher than
the maximum $T_c$, $\rho(T)$ is linear in $T$ up to about 25$\:$K.
Secondly, the exponent appears surprisingly large ($n\simeq2.7$)
at the slightly higher pressure corresponding to the maximum
$\rho_0$. This is difficult to understand without taking into
account the resistivity due to impurity scattering. In sample S1,
reported in Ref.~\onlinecite{Jaccard98}, the residual resistivity
reaches $\sim160\:\mu\Omega$cm at $P_{\rm v}$, compared to a
maximum of $35\:\mu\Omega$cm for the sample reported here.
$\rho(T)$ then showed a falloff with temperature very similar to
that of a Kondo impurity system. In other samples, this behavior
is hidden by the usual positive temperature dependence of the
resistivity. Contrary to the usual situation, where the lowest
$\rho_0$ possible is sought, this example shows how samples whose
residual resistivities are large at ambient pressure can reveal
interesting physics at high pressure. Even if a negative
temperature dependence is not seen, the power-law fit to the
resistivity is affected, deviating from the linear relationship
predicted in section \ref{sec:theory} and leading to anomalous
values of $n$. At lower pressure, the $\tilde{A}$ coefficient is
an order of magnitude larger, so (for example) almost linear
resistivity is observed at the pressure corresponding to
$T_c^{\rm{max}}$. Note that a quadratic temperature dependence of
$\rho$ was recovered at the lowest temperatures when
superconductivity was suppressed by a magnetic field greater that
$H_{c2}$.
%\\

\begin{figure}
    \includegraphics[width=\columnwidth]{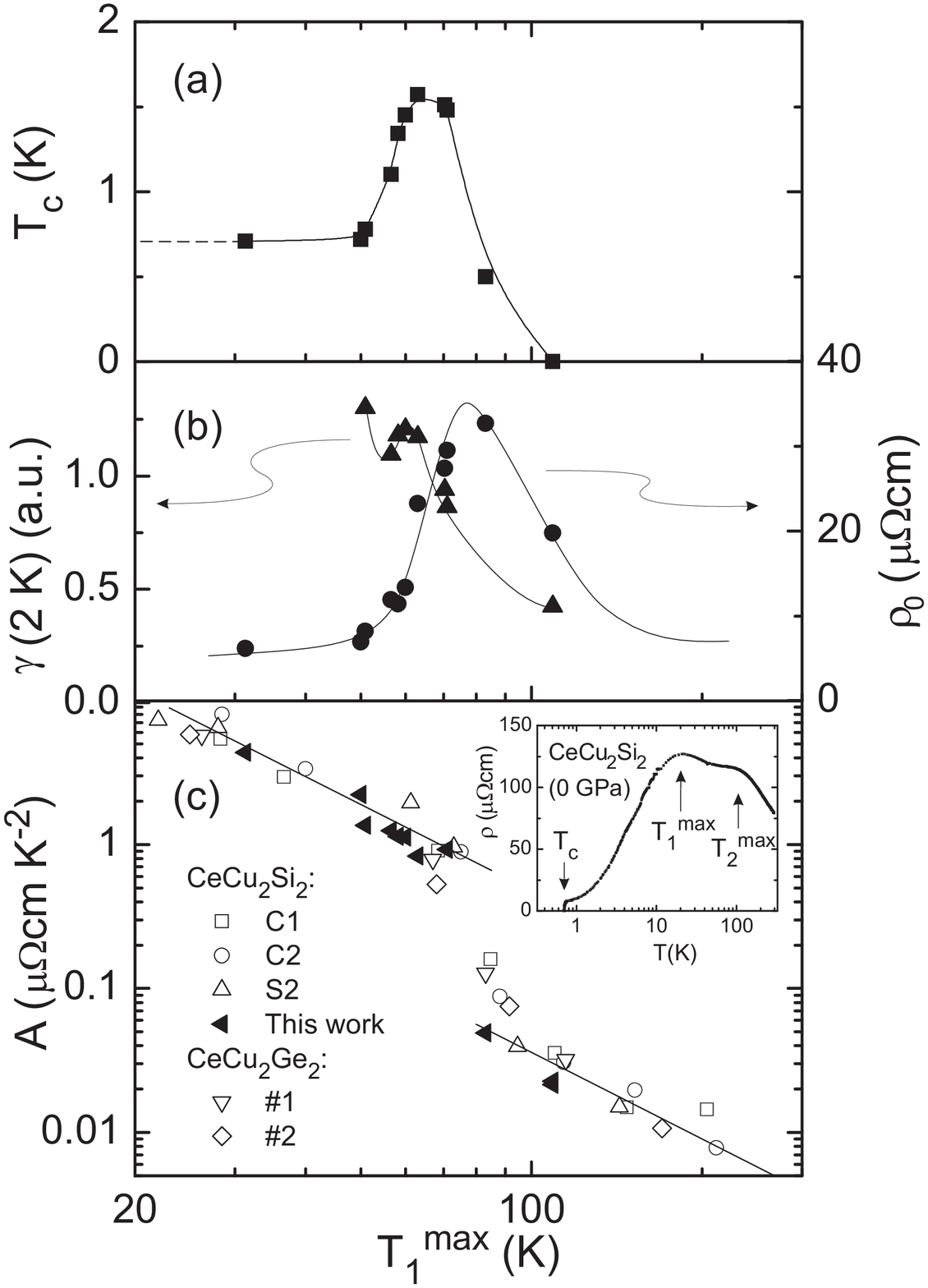}
    \caption{\label{fig:Tmax}Plotted against $T_1^{\rm{max}}$ (defined in
    inset),
     a measure of the characteristic energy scale of the system,
     are (a) the bulk superconducting transition temperature, (b) the
     residual resistivity and estimate $\tilde\gamma$ of the
     Sommerfeld coefficient, and (c) the coefficient $A$ of the
     $\rho\sim A T^2$ law of resistivity, including data from \cege.  Note the straight lines
     where the expected $A\propto (T_1^{\rm{max}})^{-2}$ scaling is
     followed.  The maximum of $T_c$ coincides with the start of
     the region where the scaling relation is broken, while the
     maximum in residual resistivity is situated in the middle of
     the collapse in $A$.  Pressure increases towards the right-hand side
     of the scale (high $T_1^{\rm{max}}$).}
\end{figure}

The normal state resistivity of heavy fermions can usually be
understood in terms of the Kondo lattice model.\cite{Cox88} At
high temperature the f-electron moments are localized and
disordered, the resistivity is large and dominated by the
scattering from spin disorder, with a characteristic $-\ln T$
slope. As the temperature is reduced, Kondo singlets form below a
characteristic temperature $T_K$, and coherence effects in a
periodic lattice cause the resistivity to drop below a maximum, at
$T_1^{\rm{max}}$, which can be considered as proportional to
$T_K$. For $T\ll T_K$ away from a critical point,
Fermi-liquid-like behavior is recovered, with $\rho \sim AT^2$,
where $A \propto T_K^{-2}$ and reflects the hugely enhanced
effective mass caused by interactions between the f-electrons.  In
a real system where $T_K$ is not too large, a second peak in the
resistivity occurs at $T_2^{\rm{max}}>T_1^{\rm{max}}$, due to the
crystalline electric field (CEF) effect\cite{Cornut72,Yamada84}
(see inset of Fig.~\ref{fig:Tmax}). The low temperature behavior
then reflects the characteristics of the lowest CEF-split f-level.
When pressure is applied, $T_2^{\rm{max}}$ remains fairly
constant, while $T_K$ rapidly increases, seen via the rise in
$T_1^{\rm{max}}$. When $T_K>\Delta_{\rm CEF}$ ($\Delta_{\rm CEF}$
is the CEF splitting between the ground and excited states) the
full 6-fold degeneracy of the J=5/2 $4\rm f^1$ multiplet is
recovered, even at the lowest temperatures. As a result the
resistivity maxima at $T_1^{\rm{max}}$ and $T_2^{\rm{max}}$ merge
into a single peak.\cite{Jaccard99} Similar behavior in the
magnetic component of the resistivity is found in all Ce compounds
studied (such as CeCu$_5$Au,\cite{Wilhelm00}
CePd$_2$Si$_2$,\cite{Demuer02} CePd$_2$Ge$_2$\cite{Wilhelm02}).
%\\

In Fig.~\ref{fig:Tmax} the $A$ vs $T_1^{\rm{max}}$ scaling is
explored in both \ce\ and \cege. The value of $A$ was determined
from the slope of the normal state resistivity versus $T^2$,
despite the non-Fermi liquid behavior shown in the inset of
Fig.~\ref{fig:rho0}. However, if one allows the exponent $n$ to
vary between 1 and 2, the resulting coefficient will not vary more
than a factor of two, which is within the scatter of the data.
There are two regions where the predicted $A \propto
(T_1^{\rm{max}})^{-2}$ relationship is followed, separated by an
abrupt drop in $A$ of over an order of magnitude. The collapse of
$A$ seems closely connected with the enhancement of
superconductivity, it is at the start of this drop that $T_c$ has
maximum, and the superconductivity has disappeared by the point
where the $A \propto (T_1^{\rm{max}})^{-2}$ scaling is recovered.
The residual resistivity however, peaks at around the midpoint of
the drop in $A$, and this is the point where $P_{\rm v}$ is
defined.

\section{Theory of T-linear resistivity and enhanced
Sommerfeld coefficient}\label{sec:theory} Various unconventional
properties observed around $P\sim P_{\rm v}$ have been explained,
at least qualitatively, by a series of theoretical investigations
on the basis of an extended Anderson lattice
model.\cite{Onishi00,Onishi002nd,Miyake02} However, the $T$-linear
temperature dependence of the resistivity observed in a narrow
region around $P\sim P_{\rm v}$ remains as yet unexplained.  In
Ref.~\onlinecite{Onishi00}, microscopic calculations showed that
the static limit of the effective interaction $\Gamma^{(0)}(q)$
between quasiparticles is enhanced greatly around $P\sim P_{\rm
v}$, and is almost independent of $q$, the momentum transfer, up
to $\sim$3/2 of $p_{\rm F}$, reflecting the local nature of
critical valence fluctuations.  This implies that the valence
fluctuation response function $\chi_{\rm v}(q,\omega)$, is also
almost $q$-independent in the low frequency region.  Based on this
observation, we present here a phenomenological theory explaining
the $T$-linear resistivity and the enhancement of the Sommerfeld
coefficient $\gamma$ around $P\sim P_{\rm v}$.
%\\

We adopt an exponentially decaying phenomenological form for the
valence-fluctuation propagator (dynamical valence susceptibility)
$\chi_{\rm v}$:
\begin{eqnarray}
\chi_{\rm v}(q,\omega)&\equiv& {\rm i}\int_{0}^{\infty}{\rm d}t
e^{{\rm i}\omega t} \langle[n_{\rm f}(q,t),n_{\rm f}(-q,0)]\rangle
\\
&=&{K\over \omega_{\rm v}-{\rm i}\omega}, \quad{\hbox{for
$q<q_{\rm c}\sim p_{\rm F}$}} \label{chiv1}
\end{eqnarray}
where $n_{\rm f}(q)$ is the Fourier component of the number of
f-electron per Ce site, $K$ is a constant of ${\cal O}(1)$, and
$\omega_{\rm v}$ parameterizes the closeness to criticality.
$\omega_{\rm v}$ is inversely proportional to the valence
susceptibility $\chi_{\rm v}(0,0)= -(\partial n_{\rm f}/\partial
\epsilon_{\rm f})_{\mu}$.
%\\

\begin{figure}
\includegraphics[width=0.75\columnwidth]{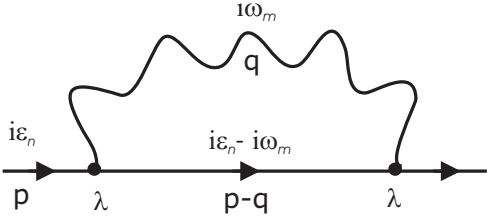}
%$\feyn{fsfglffs}$
\caption{\label{fig:feyn} Feynman diagram for the self-energy
given by eqs.~(\ref{chiv2}) and (\ref{chiv3}). The solid line
represents the Green function of the quasiparticles, the wavy line
the propagator of the valence fluctuations, and the filled circle
the coupling between valence-fluctuation modes and the
quasiparticles. $\epsilon_n$ and $\omega_m$ are the Matsubara
frequency of the quasiparticle and fluctuation propagators,
respectively.}
\end{figure}

The real and imaginary parts of the retarded self-energy
$\Sigma_{\rm vf}^{\rm R}(p,\epsilon+{\rm i}\delta)$ respectively
give a measure of the quasiparticle effective mass and lifetime.
They can be calculated using a simple one-fluctuation mode
exchange process (see Fig.~\ref{fig:feyn}) and given as follows:
\begin{eqnarray}
{\rm Re}\Sigma_{\rm vf}^{\rm R}(p,\epsilon)&=& -{K\over
2\pi}\sum_{{\bf q}}|\lambda|^{2} \int_{-\infty}^{+\infty}{\rm
d}x{x\over \omega_{\rm v}^{2}+x^{2}} \nonumber
\\
& &\qquad\qquad\times {\coth{x\over 2T}+\tanh{\xi_{{\bf p}-{\bf
q}}\over 2T}\over -\epsilon+\xi_{{\bf p}-{\bf q}}+x},
\label{chiv2}
\end{eqnarray}
\begin{eqnarray}
{\rm Im}\Sigma_{\rm vf}^{\rm R}(p,\epsilon) &=&-{K\over
2}\sum_{{\bf q}}|\lambda|^{2} {\epsilon-\xi_{{\bf p}-{\bf
q}}\over\omega_{\rm v}^{2}+ (\epsilon-\xi_{{\bf p}-{\bf q}})^{2}}
\nonumber
\\
& &\times \biggl(\coth{\epsilon-\xi_{{\bf p}-{\bf q}}\over 2T}
+\tanh{\xi_{{\bf p}-{\bf q}}\over 2T}\biggr), \label{chiv3}
\end{eqnarray}
where $\lambda$ is the coupling between quasiparticles and the
valence fluctuation modes, and $\xi_{p}$ is the dispersion of the
quasiparticle. For simplicity, $\lambda$ is assumed to be constant
without wavenumber or frequency dependence.
%\vskip5cm \hskip6cm Fig.*
%\\
%\\

In typical limiting cases, (\ref{chiv3}) can be straightforwardly
calculated in the approximation $\xi_{{\bf p}-{\bf q}}\simeq
-vq\cos\theta$, where $\theta$ is the angle between ${\bf p}$ and
${\bf q}$, $v$ is the quasiparticle velocity, and $p$ is
assumed to be on the Fermi surface, i.e., $p=p_{\rm F}$:  \\
\\
$T=0$, $\epsilon\not=0$:
\begin{equation}
{\rm Im}\Sigma_{\rm vf}^{\rm R}( p_{\rm F} ,\epsilon)\simeq
-{|\lambda|^{2}Kq_{\rm c}^{2}\over32\pi^{2} v}
\ln\biggl(1+{\epsilon^{2}\over\omega_{\rm v}^{2}}\biggr),
\label{chiv4}
\end{equation}
where $q_{\rm c}$ is the cutoff wavenumber of the order of $k_{\rm
F}$.
\\
 \\
$\epsilon=0$, $0<T\ll\epsilon_{\rm F}$:
\begin{eqnarray}
{\rm Im}\Sigma_{\rm vf}^{\rm R}(p_{\rm F},0)&\simeq&
-{|\lambda|^{2}K\over8\pi^{2} v} \int_{0}^{q_{\rm c}}{\rm d}q\,q
\int_{-vq/2T}^{vq/2T}{\rm d}
y \times\nonumber\\
& & { y \over (\omega_{\rm v}/T)^{2}+ y^{2}}(\coth{ y \over
2}-\tanh{ y \over2}) \label{chiv5},
\end{eqnarray}
where $y=vq\cos\theta/2T$. Since $vq\gg T$ holds in the dominant
region of q-space, the integration with respect to $y$ can be
performed, to a good accuracy, leading to
\begin{equation}
{\rm Im}\Sigma_{\rm vf}^{\rm R}( p_{\rm F},0)\simeq
-{|\lambda|^{2}Kq_{\rm c}^{2}\over4\pi^{2} v} {T\over \omega_{\rm
v}}\tan^{-1}{T\over \omega_{\rm v}}, \label{chiv5a}
\end{equation}
where we have made approximation that the range of integration is
restricted as $-1<y<1$ in which the last factor in (\ref{chiv5})
is approximated as $2/y$.  Then,
\begin{equation}
{\rm Im}\Sigma_{\rm vf}^{\rm R}(p_{\rm F},0)\simeq
-{|\lambda|^{2}Kq_{\rm c}^{2}\over4\pi^{2} v}
\cases{\biggl({T\over \omega_{\rm v}}\biggr)^{2},&$T\ll\omega_{\rm
v}$\cr {\pi\over 2}{T\over \omega_{\rm v}},&$T\gg\omega_{\rm
v}$\cr} \label{chiv5b}
\end{equation}
The latter result, ${\rm Im}\Sigma_{\rm vf}(p_{\rm
F},\epsilon=0)\propto T/\omega_{\rm v}$ for $T\gg\omega_{\rm v}$,
implies that almost all the critical valence-fluctuation modes can
be regarded as classical at $T>\omega_{\rm v}$, and $T$-linear
dependence stems from the asymptotic form of $\coth(x/2T)\simeq
2T/x$, essentially the classical approximation of the Bose
distribution function.
%\\

The real part of the self-energy, (\ref{chiv2}), can be calculated
easily at $T=0$ and $\epsilon\sim 0$, leading to
    \begin{eqnarray}
{\rm Re}\Sigma_{\rm vf}^{\rm R}(p_{\rm F},\epsilon) &-&\Sigma_{\rm
vf}^{\rm R}(p_{\rm F},0)\nonumber
\\
&\simeq& -{|\lambda|^{2}K\epsilon\over4\pi^{2}} \int_{0}^{q_{\rm
c}}{\rm d}q\,q^{2}\int_{-1}^{1}{\rm d}t \nonumber
\\
& &\times \biggl[ {-1\over\omega_{\rm
v}^{2}+(vqt)^{2}}\ln\biggl|{e\omega_{\rm v}\over vqt}\biggr|
+{\pi\omega_{\rm v} vq|t|\over[\omega_{\rm v}^{2}+(vqt)^{2}]^{2}}
\nonumber
\\
& &\quad\qquad +{2\omega_{\rm v}^{2}\over[\omega_{\rm
v}^{2}+(vqt)^{2}]^{2}}\ln\biggl|{\omega_{\rm v}\over vqt}\biggr|
\biggr], \label{chiv6}
\end{eqnarray}
where $t=\cos\theta$. In the limit $\omega_{\rm v}\ll v p_{\rm
F}$, integration with respect to $t$ in (\ref{chiv6}) leads to
\begin{eqnarray}
{\rm Re}\Sigma_{\rm vf}^{\rm R}(p_{\rm F},\epsilon)&\simeq&
-{|\lambda|^{2}Kq_{\rm c}^{2}\over2\pi^{2}
v}{\epsilon\over\omega_{\rm v}} \int_{0}^{1}{\rm d} u
{1-u^{2}\over u^{2}+1} \ln\biggl|{1\over u }\biggr| \label{chiv7}
\\
&\propto&-{\epsilon\over \omega_{\rm v}}, \label{chiv7a}
\end{eqnarray}
where $u=vqt/\omega_{\rm v}$.
%\\

The $T$-linear dependence of ${\rm Im}\Sigma_{\rm vf}^{\rm
R}(p,0)$, for $T>\omega_{\rm v}$, (\ref{chiv5b}), implies
$T$-linear resistivity, as the quasiparticles are subject to the
large angle scattering by the critical valence-fluctuation modes.
These are effective in a wide region in the Brillouin zone due to
their local nature and easily couple to the Umklapp process of
quasiparticle scattering.  This result is consistent with the
experimental fact that $T$-linear resistivity is observed in a
narrow pressure region around $P_{\rm v}$, which is considered to
correspond to a nearly critical valence transition of the Ce ion.
%\\

Such a $T$-linear dependence has been discussed in the context of
high-$T_c$ cuprates with a marginal Fermi liquid (MFL)
assumption,\cite{Varma1} and charge transfer fluctuations were
once considered as an origin for MFL,\cite{Varma2,Varma2a} while
further theoretical models have been put forth up to
now.\cite{Varma3} Excepting the $T$-linear resistivity, the
present result is different from MFL behavior. The self-energy
exhibits different energy dependence, while the idea for the
origin of our singular behavior shares aspects similar to the
first idea of a charge transfer mechanism for high-$T_c$
cuprates.\cite{Varma2,Varma2a} $\Sigma(\epsilon)$ in the MFL model
is given as $\Sigma(\epsilon)\propto(\epsilon\ln\epsilon-{\rm
i}|\epsilon|)$ \cite{Varma1} which is indeed different from the
present case [Eqs.(\ref{chiv4}) and (\ref{chiv7a})]. In any case,
it is to be noted that $T$-linear resistivity is accompanied by
the peak of $T_c$ in both systems, high-$T_c$ cuprates and
CeCu$_2$Si$_2$.
%\\

The result (\ref{chiv7a}) implies that the mass enhancement
$(1-\partial{\rm Re}\Sigma^{\rm R}_{\rm
vf}(\epsilon)/\partial\epsilon)$ is expected around $P\sim P_{\rm
v}$. Namely, the effective mass is given by
\begin{equation}
m^{*}\propto {\bar m}{1\over\omega_{\rm v}}, \label{chiv8}
\end{equation}
where ${\bar m}$ is the effective mass renormalized by the
conventional correlation effect, leading to heavy electrons, i.e.
not including the effect of critical valence fluctuations. This
latter effective mass ${\bar m}$ exhibits a drastic decrease
around $P\sim P_{\rm v}$, while the second factor in (\ref{chiv8})
is enhanced.  Both effects should be reflected in the Sommerfeld
coefficient $\gamma$, so that the peak of $\gamma \propto m^*$ is
shifted to the lower pressure (larger $\bar m$) side, and the
anomaly of $\gamma$ due to the valence fluctuations may be smeared
to some extent.  Nevertheless, some trace should be observed. (The
shift of peak of $\gamma$ can be understood as the superposition
of the two trends using a model $P$-dependence of ${\bar m}$ and
$\omega_{\rm v}$.) Indeed, the present experimental result
presented in Figs.~\ref{fig:gamma} and~\ref{fig:Tmax} may be
explained by this effect.
%\\
\section{Discussion}
Our calorimetric results in such extreme conditions deserve some
discussion, in particular the considerable apparent increase in
the specific heat jump at the superconducting transition when
$P_{\rm v}$ is approached. A very large specific heat jump at
$T_c$ would be strongly reminiscent of the huge value found in
CeCoIn$_5$.\cite{Petrovic01}  It is therefore a legitimate
question to ask how much the results of the uncalibrated AC
calorimetry technique under pressure can be relied on to give an
accurate measurement of the specific heat.
%\\

The model used to extract the specific heat from the amplitude and
phase of the temperature oscillations takes no account of the heat
capacity of the solid helium, diamonds, or surrounding pressure
apparatus, or the essentially three-dimensional nature of the
situation. Secondly, the thermopower of the \underline{Au}Fe
thermocouple has been assumed not to vary with pressure
(Ref.~\onlinecite{Wilhelm02} indicates that it varies by no more
than 20\% up to 12 GPa).
%\\

Nevertheless, the superconducting transition observed corresponds
to $\sim100\%$ of the signal amplitude, indicating that the
addenda are a minority contribution to the total signal. Runs at
several different frequencies agree to within 10--20\% after the
amplitude and phase are combined, with the discrepancy possibly
due to frequency-dependent addenda. Kapitza resistance between the
sample and helium is likely to better decouple the sample from its
surroundings at very low temperature. If the specific heat is
calculated using the two-frequency method [eq.~(\ref{eq:Tdc})],
the result agrees ($<5\%$) with that calculated using the
amplitude and phase up to at least $2T_c$. Given these
observations, it seems reasonable to accept our results as a good
first approximation to $C_P$, to within a constant scaling factor,
and with an unknown but relatively small component due to addenda.
%\\

Furthermore, the apparent anomaly in the normal state specific
heat shown in Fig.~\ref{fig:gamma} was measured at a fixed
temperature and frequency above the superconducting transition,
with pressure the only independent variable.  The small peak in
$\gamma$ is consistent with the maximum in the initial slope of
the upper critical field observed at the same pressure, though the
interpretation of the latter depends on whether the sample can be
considered to be in the clean or dirty limit, or somewhere in
between.
%\\

Having addressed the experimental questions, let us discuss some
other remaining points. The merging of $T_{1}^{\rm max}$ and
$T_{2}^{\rm max}$ seems to be be a general feature at $P_{\rm v}$
in compounds where a critical valence transition is thought to
exist. It can be understood as follows:

The so-called Kondo temperature $T_{\rm K}$, related to
$T_{i}^{\rm max}$ ($i=1,2$), depends crucially on the degeneracy
$(2\ell+1)$ of the local f-state: $T_{\rm K}\sim
D\exp[-1/(2\ell+1)\rho_{\rm F}|J|]$, where $D$ is the bandwidth of
conduction electrons, $\rho_{\rm F}$ the density of states of
conduction electrons at the Fermi level, and $J$ the c-f exchange
coupling constant.\cite{Okada73} Even though the 6-fold degeneracy
of the 4f-state is lifted by the CEF effect, leaving the Kramers
doublet ground state and excited CEF levels with excitation energy
$\Delta_{\rm CEF}$, the Kondo temperature $T_{\rm K}$ is still
enhanced considerably by the effect of the excited CEF
levels.\cite{Yamada84}
%\\

The technical degeneracy relevant to the Kondo effect is affected
by the broadening $\Delta E$ of the lowest CEF level. If $\Delta
E\ll\Delta_{\rm CEF}$, the degeneracy relevant to $T_{\rm K}$ is
2-fold.  On the other hand, if $\Delta E>\Delta_{\rm CEF}$, it
increases to 4- or 6-fold.  The level broadening is given by
$\Delta E\simeq z\pi\rho_{\rm F}|V|^{2}$ where $|V|$ is the
strength of c-f hybridization, and $z$ is the renormalization
factor which gives the inverse of mass enhancement in the case of
a lattice system. It is crucial that $\Delta E$ is very sensitive
to the valence of Ce ion because $z$ is essentially given by $q$
[eq.(\ref{eq:m*nf})]. In particular, the factor $z$ increases from
a tiny value in the Kondo regime, $z\sim(1-n_{\rm f})\ll 1$, and
approaches unity in the so-called valence fluctuation regime.
%\\

Since the factor $\pi\rho_{\rm F}|V|^{2}\gg\Delta_{\rm CEF}$ in
general for Ce-based heavy electron systems, the ratio $\Delta
E/\Delta_{\rm CEF}$, which is much smaller than 1 in the Kondo
regime, greatly exceeds 1 across the valence transformation around
$P\sim P_{\rm v}$, leading to the increase of the technical
degeneracy of f-state, {\it irrespective} of the sharpness of the
valence transformation. Therefore, $T_{1}^{\rm max}$ should merge
with $T_{2}^{\rm max}$, which corresponds to 4- or 6-fold
degeneracy of 4f-state due to the effect of finite temperature,
i.e., $T\sim \Delta_{\rm CEF}$.  This may be the reason why
$T_{1}^{\rm max}$ increases and approaches $T_{2}^{\rm max}$ at
pressure where $T_c$ exhibits the maximum, and the KW ratio
changes between strongly and weakly correlated classes.
%\\

While the experimental picture of \ce~presented in this paper is
more complete than the theoretical, a large number of the features
found around $P_{\rm v}$ follow directly from the valence
fluctuation approach and the addition of a $U_{\rm cf}$ term to
the hamiltonian. The linear resistivity is explained in section
\ref{sec:theory}, as is the local maximum in the electronic
specific heat, possibly due to the renormalisation of the
effective mass due to valence fluctuations, superimposed on an
overall decrease with pressure. The enhancement of the residual
resistivity at low temperature follows from the renormalisation of
impurity potentials by valence fluctuations. The relative
positions of the peaks in $T_c$, $\gamma$, and $\rho_0$ are
consistent with the valence fluctuation scenario, but for a more
precise comparison more detailed calculation would be needed.
%\\

Other features yet to be fully addressed with the current model
are observed to occur in the valence fluctuation region. They are
the apparent increase in the specific heat jump at $T_c$, the
temperature dependence of the impurity contribution to the
resistivity, and the nature of the superconducting state between
the onset and completion of the superconducting transition.
%\\

The presence, and indeed enhancement, of superconductivity so far
from  the disappearance of magnetic order calls into question
whether magnetic mediation is really the sole mechanism of
superconductivity in \ce. The evidence presented here, along with
other anomalous behavior seen at a pressure well separated from
the disappearance of magnetism, strongly suggests the presence of
a second quantum critical point in \ce, this time related to
quantum fluctuations between electronic configurations rather than
to collective spin instabilities. While magnetic pairing may be
responsible for superconductivity at the magnetic QCP, critical
valence fluctuations are responsible for pairing at $P_{\rm v}$.
The recent result in
CeCu$_2$(Si$_{0.9}$Ge$_{0.1}$)$_2$,\cite{Yuan03} where two
separate peaks of $T_c$ are observed, suggests the validity of the
present point of view.
 \begin{figure}
 \includegraphics[width=0.9\columnwidth]{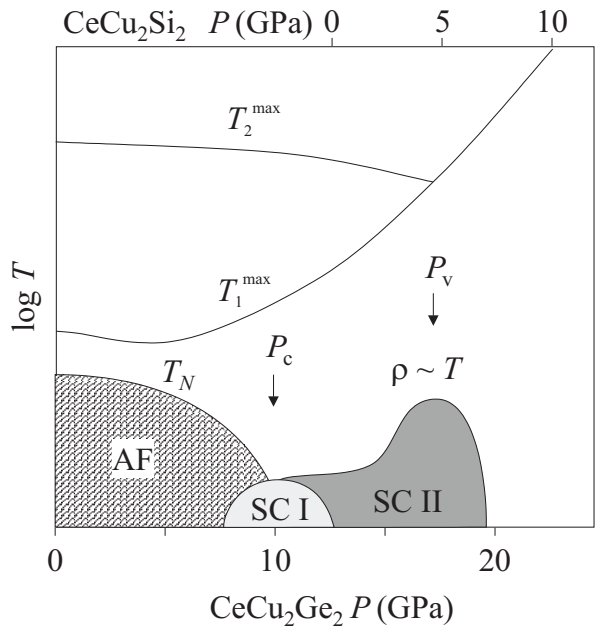}
 \caption{\label{fig:schemphadiag} Schematic $P$-$T$ phase diagram for
CeCu$_2$(Si/Ge)$_2$ showing the two critical pressures
 $P_{\rm c}$ and $P_{\rm v}$. At $P_{\rm c}$, where the antiferromagnetic ordering
temperature $T_N\rightarrow 0$,
  superconductivity in region $SC\ I$ is mediated by
 antiferromagnetic spin fluctuations; around $P_{\rm v}$, in the region $SC\
II$, valence
 fluctuations provide the pairing
 mechanism and the resistivity is linear in temperature.  The
 Temperatures $T_1^{\rm{max}}$, and $T_1^{\rm{max}}$,
 merge at a pressure coinciding with $P_{\rm v}$.}
 \end{figure}

Figure~\ref{fig:schemphadiag} shows a schematic phase diagram for
the CeCu$_2$(Si/Ge)$_2$ system.  The two critical pressures,
$P_{\rm c}$ and $P_{\rm v}$, are respectively defined by the
disappearance of magnetic order as $T_N$ tends to zero, and by the
region of linear resistivity where $\rho_0$ has a maximum and
$T_1^{\rm{max}}\simeq T_2^{\rm{max}}$, accompanied by a maximum in
$T_c$. In \ce\ and \cege\ the two critical pressures are widely
separated. In compounds such as CePd$_2$Si$_2$ on the other hand,
superconductivity is found in a narrow pocket, seemingly directly
connected to the disappearance of magnetism as $T_N\rightarrow 0$.
However, many of the other anomalies listed in table
\ref{tab:properties} are still observed in this system, and are
difficult to explain within a purely spin fluctuation picture.  If
a valence instability is present in CePd$_2$Si$_2$, $P_{\rm v}$ is
superimposed on $P_{\rm c}$, as identified by the pressure at
which $T_1^{\rm{max}}\simeq T_2^{\rm{max}}$.\cite{Demuer02} The
physics associated with valence change in \ce~may thus also play
an important role in other heavy fermion superconductors. Linear
resistivity and an enhancement of $\rho_0$ have also been seen in
the CeTIn$_5$ compounds,\cite{Muramatsu01,Petrovic01} where T is
Co, Rh or Ir. For this family, superconductivity extends over a
relatively broad pressure range, and it may be that valence
fluctuations also play a role with a critical valence pressure
separate from any magnetic instability.
%\\

Finally, it is worth addressing the physical interpretation of the
valence fluctuation mediated pairing interaction. We emphasize
that this intuitive explanation is rather speculative, but we
think that it is sufficiently useful to merit inclusion.
%\\

A clue comes from the likely nearest neighbor pairing, implied by
the largely local nature of the interaction, and the prediction of
$d$-wave pairing symmetry. One can imagine an almost filled
f-band, with each occupied $\rm f^1$ site experiencing a Coulomb
repulsion $U_{\rm cf}$ from the respective conduction electrons.
As the pressure is increased and $\epsilon_{\rm f}$ moves closer
to the Fermi level $\epsilon_{\rm F}$, there will come a point
where $\epsilon_{\rm f}+U_{\rm cf}=\epsilon_{\rm F}$ and the
f-band will start to empty.  On an individual $4\rm f^0$ `hole'
site, the $U_{\rm cf}$ interaction will be absent, thus an
increased density of conduction electrons would be energetically
favorable at this position. If this extra `screening' conduction
electron density is not strictly localized onto the atom itself,
but spills onto neighboring sites, the f-electrons on Ce atoms
around the original `hole' site will feel an increased repulsion.
The tendency to transfer electrons from the f to conduction bands
will be locally reinforced, explaining intuitively the
increasingly catastrophic drop in $n_{\rm f}$ for larger $U_{\rm
cf}$, predicted in Ref.~\onlinecite{Onishi00}. For large enough
$U_{\rm cf}$, phase separation would be expected to occur for some
values of $\epsilon_{\rm f}$.
%\\

The attractive pairing interaction can be understood as follows:
Consider an isolated pair of $4\rm f^0$ `holes', accompanied by
their cloud of conduction electrons.  If these are separated by
two lattice positions, with an intervening filled $4\rm f^1$ site,
the two clouds of conduction electrons will overlap at the
intermediate site, further increasing the Coulomb energy at that
point. It would therefore be energetically favorable for the two
`holes' to be on neighboring atoms, thus the attractive
interaction. The attractive interaction between `holes' is
equivalent to that between `electrons', so that this argument
would give an intuitive understanding of the origin of the
valence-fluctuation mechanism of superconductivity.
%\\

\section{Conclusions}
The enhancement of superconductivity in \ce~under pressure is
found to coincide with a number of anomalies in the
superconducting and normal state properties that are hard to
explain in a purely spin fluctuation scenario. Many of these
anomalies are directly related to an abrupt change in valence of
the Ce ion, while others can be indirectly connected to such a
transition. We propose a second critical pressure $P_{\rm v}$ at
around 4.5$\:$GPa where critical valence fluctuations provide the
superconducting pairing mechanism. An extended Anderson lattice
model with Coulomb repulsion between the conduction and
f-electrons predicts an abrupt change in Ce f-level occupation.
The associated fluctuations are sufficient to explain the observed
enhancement of $T_c$, the $T$-linear normal state resistivity, the
enhancement of the residual resistivity, and the peak in the
electronic specific heat coefficient $\gamma$.

\begin{acknowledgments}
A.H. and D.J. would like to thank A. Demuer, F. Bouquet and A.
Junod for careful reading of the manuscript and invaluable
comments, and R. Cartoni for technical assistance. One of the
authors (K.M.) acknowledges stimulating communications with H.Q.
Yuan and F. Steglich.  K.M. was supported in part by the COE
program (No.10CE2004) by MEXT of Japan.
\end{acknowledgments}

\end{document}